\documentclass[pra,aps,showpacs,groupedaddress,superscriptaddress,twocolumn]{revtex4-1}
\usepackage{amsmath,amssymb}
\usepackage[utf8x]{inputenc}
\usepackage[english]{babel}
\usepackage{graphicx,epsfig}
\usepackage{todonotes}
\usepackage{hyperref}

 \global\long\def\abs#1{\left|#1\right|}

\usepackage{braket}
\usepackage{soul}
\usepackage{color}
\makeatother
\usepackage{placeins}

\begin{document}

\title{Correlation engineering via non-local dissipation}

\author{K. Seetharam}
\affiliation{Department of Electrical Engineering, Massachusetts Institute of Technology, Cambridge, Massachusetts 02139, USA}
\affiliation{Department of Physics, Harvard University, Cambridge MA, 02138, USA}
\author{A. Lerose}
\affiliation{Department of Theoretical Physics, University of Geneva, 1211 Geneva, Switzerland}
\author{R. Fazio}
\affiliation{International Center for Theoretical Physics ICTP, Strada Costiera 11, I-34151, Trieste, Italy}
\affiliation{Dipartimento di Fisica, Universita di Napoli 'Federico II', Monte S. Angelo, I-80126 Napoli, Italy}
\author{J. Marino}
\affiliation{Department of Physics, Harvard University, Cambridge MA, 02138, USA}
\affiliation{Institut f\"ur Physik, Johannes Gutenberg Universit\"at Mainz, D-55099 Mainz, Germany}


\begin{abstract}


Controlling the spread of correlations in quantum many-body systems is a key challenge at the heart of quantum science and technology. Correlations are usually destroyed by dissipation arising from coupling between a system and its environment. Here, we show that dissipation can instead be used to engineer a wide variety of spatio-temporal correlation profiles in an easily tunable manner. We describe how dissipation with any translationally-invariant spatial profile can be realized in cold atoms trapped in an optical cavity. A uniform external field and the choice of spatial profile can be used to design when and how dissipation creates or destroys correlations. We demonstrate this control by generating entanglement preferentially sensitive to a desired spatial component of a magnetic field. We thus establish non-local dissipation as a new route towards engineering the far-from-equilibrium dynamics of quantum information, with potential applications in quantum metrology, state preparation, and transport.



\end{abstract}


\maketitle


Correlations in many-body systems allow us to monitor the dynamics of quantum information by giving insight, for example, into the growth of quantum fluctuations and entanglement~\cite{cheneau2012light,calabrese2006time,islam2015measuring}. While dissipation usually decoheres the system, it can also be used to prepare correlated quantum states that are a powerful resource for quantum information processing~\cite{diehl2008quantum,eisert2010noise,verstraete2009quantum,diehl2011topology,  PhysRevA.86.013606,buvca2019non}. Compared to the conventional use of unitary processes to manipulate a system, the irreversiblity of dissipative dynamics makes it more robust to variations in the initial state and allows for simpler control protocols.



Realizing the potential of dissipation engineering has been challenging, with experiments thus far using a combination of unitary operations and dissipation to produce and stabilize entangled states of a small number of qubits~\cite{barreiro2011open,lin2013dissipative,ma2019dissipatively}. Purely non-unitary preparation of correlated states typically requires dissipation that is non-local in space and can lock the phases of two or more adjacent particles~\cite{diehl2008quantum}. Correlations generated by such dissipation, even with spatial profiles involving only neighboring particles, can endow a system with exotic character such as non-trivial topological properties~\cite{diehl2011topology,bardyn2013topology,bardyn2012majorana,iemini2015localized, tonielli2020topological}, quantum critical points without equilibrium counterparts~\cite{eisert2010noise,hoening2012critical,marcos2012photon, marino2016driven}, and integrability revival in the presence of a drive~\cite{lange2018time,reiter2019engineering}. Experimental implementations of non-local dissipation with long-range spatial profiles have been proposed in atomic platforms~\cite{PhysRevA.89.023616,parmee2018phases}, but with limited tunability of the profile, and thereby of the generated correlations and accessible effects. For example, dissipation with a power-law spatial profile may enable the realization of exotic phenomena such as purely
non-unitary many-body quantum synchronization~\cite{buvca2019dissipation,buca2021algebraic} or novel non-equilibrium critical states that would be otherwise inaccessible~\cite{dogra2019dissipation}. The ability to easily tune the spatial profile of dissipative channels would therefore open new avenues and applications of dissipation engineering.



Our work demonstrates a practical route towards dissipative quantum information processing, showing that dissipation with a fully customizable spatial profile is readily realizable in systems of cold atoms trapped in a single-mode cavity, and furthermore that the behavior of this dissipative channel can be modulated via a uniform external field. Control over the spatial profile and uniform field can be exploited to engineer the profile of correlations in the system, which we show by tailoring the spatio-temporal window over which correlations are present, creating oscillating packets of correlations, and sending the system towards an increasingly squeezed state. The ability to shape correlations enables the manipulation of entanglement dynamics, which we demonstrate by generating entanglement that preferentially enhances metrological sensitivity to a desired spatial mode of an external field.

\begin{figure}[t!]
	\centering
	\includegraphics[clip, width=0.85\columnwidth]{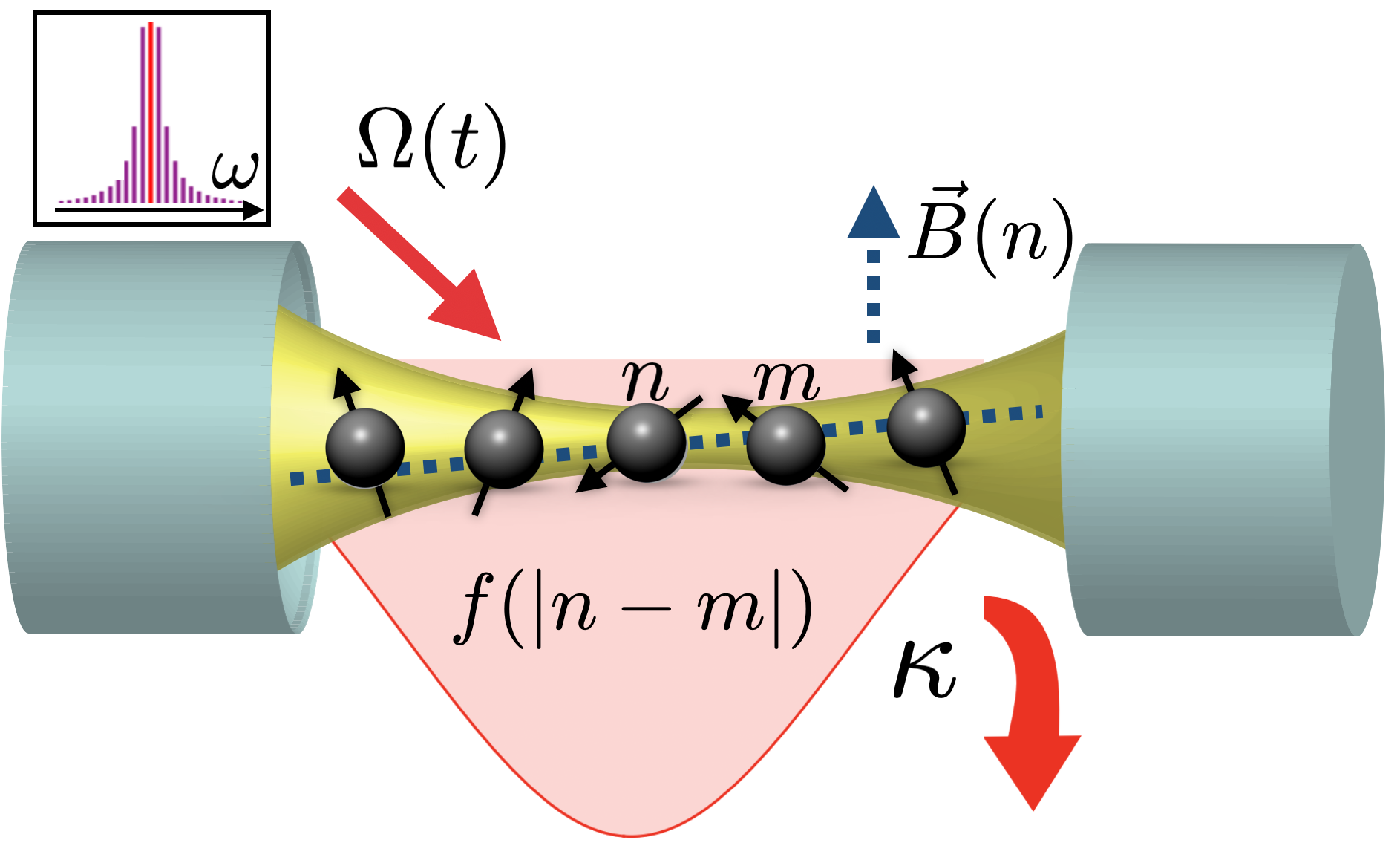}
	\caption{\textbf{Experimental realization.} Spin degrees of freedom are encoded in the internal states of atoms trapped in a leaky optical cavity. A magnetic field gradient, $\vec{B}(n)$, and a classical Raman beam, $\Omega(t)$, with multiple sidebands (inset) are used to generate a desired spatial profile, $f(|n-m|)$, of non-local dissipation.}
	\label{fig:ExpSchematic}
\end{figure}

We first illustrate the novel dynamics enabled by non-local dissipation. Consider a translationally-invariant, one-dimensional many-body quantum system undergoing both unitary dynamics and Markovian dissipation. The state of the system, $\rho$, evolves according to the quantum master equation in Lindblad form
\begin{equation}\label{eq:model}
	\resizebox{.896\hsize}{!}{$	\dot{\rho}=i\left[\rho,\hat{H}\right]
	+\kappa\sum_{n,m}f_{n,m}\left(\hat{L}_{n}\rho\hat{L}_{m}^{\dagger}-\frac{1}{2}\left\{ \hat{L}_{m}^{\dagger}\hat{L}_{n},\rho\right\} \right)$},
\end{equation}
where $\hat{H}$ is the Hamiltonian characterizing unitary evolution, $\hat{L}_{n}$ is the jump operator characterizing the loss channel, and $n,m=1...N$ index the sites of the chain. Here, $f_{n,m}$ is the spatial profile of the dissipation and only depends on the difference $\left|n-m\right|$. Independent dissipation, corresponding to $f_{n,m}=\delta_{n,m}$, and collective dissipation, corresponding to $f_{n,m}=1$, are the two commonly considered scenarios. The former is a common source of decoherence in experiments, while the latter can generate collective entanglement useful for quantum metrology~\cite{morrison2008dynamical,PhysRevA.66.022314, PhysRevLett.107.080503,kessler2012dissipative,PhysRevLett.110.120402,reiter2016scalable, floren}. Both these loss channels are spatially homogeneous and therefore cannot cause correlations to spread in space.

The case of tunably non-local dissipation can be understood as interpolating between independent and collective loss. For example, consider a short-range spatial profile, $f_{n,m}=e^{-|n-m|/\chi}$, where $\chi$ is the length scale of the profile. If the system is comprised of atoms coupled to a common cavity  mode, with dissipation arising from photons leaking out of the cavity, detection of a leaked photon does not allow one to discern which specific atom emitted the photon. Instead, such a photon can only be traced back to a neighborhood of atoms comprised of approximately $\chi$ sites. As $\chi$ is decreased or increased, we recover independent and collective dissipation respectively.

\begin{figure*}[t!]
	\centering
	\includegraphics[width=0.90\textwidth, height=0.38\textheight]{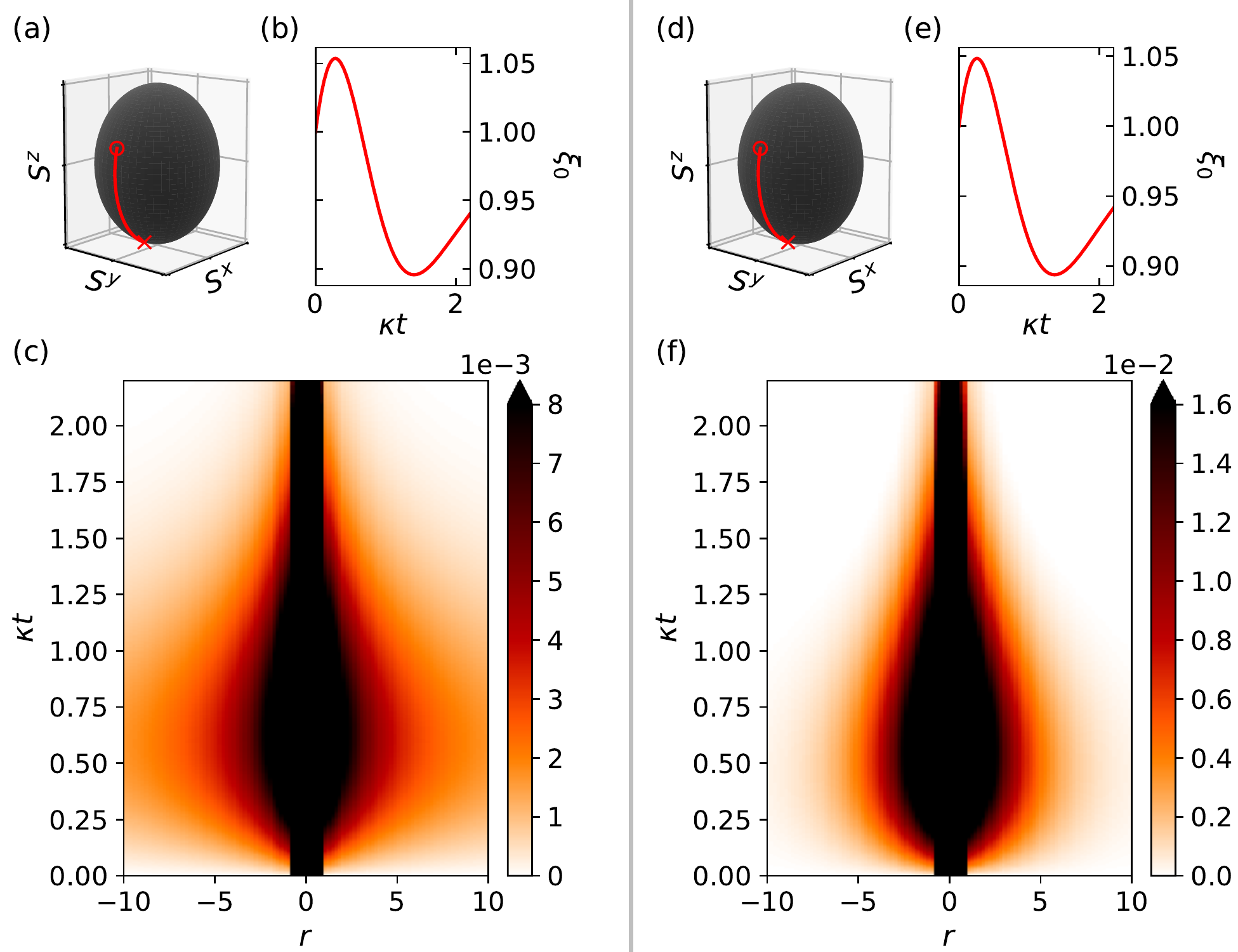}
	\caption{\textbf{Dynamical confinement via dissipation.} Panels (a)-(c) correspond to a long-range spatial profile with $\alpha=1.25$. Panels (d)-(f) correspond to a short-range spatial profile with $\chi=2.0$. For both cases, we choose $\kappa=1$ and initialize the system in a spin coherent state pointing in the direction $\theta(t=0)=0.4\pi$, $\phi(t=0)=0$. \textbf{(a), (d)} Motion of the collective spin on the Bloch sphere. \textbf{(b), (e)} Squeezing of the collective spin. \textbf{(c), (f)} Connected correlation function $C^{zz}\left(r,t\right)$. Correlations spread and then contract in accordance to the motion of the collective spin. Upper bounds are imposed on the color scales to visually highlight correlation profiles both here and in Fig.~\ref{fig:CorrPatterns};  adjacent sites numerically take on values of order $10^{-2}$ to $10^{-1}$}.
	\label{fig:DispConfinement}
\end{figure*}

\begin{figure*}[t!]
	\centering
	\includegraphics[width=0.90\textwidth, height=0.43\textheight]{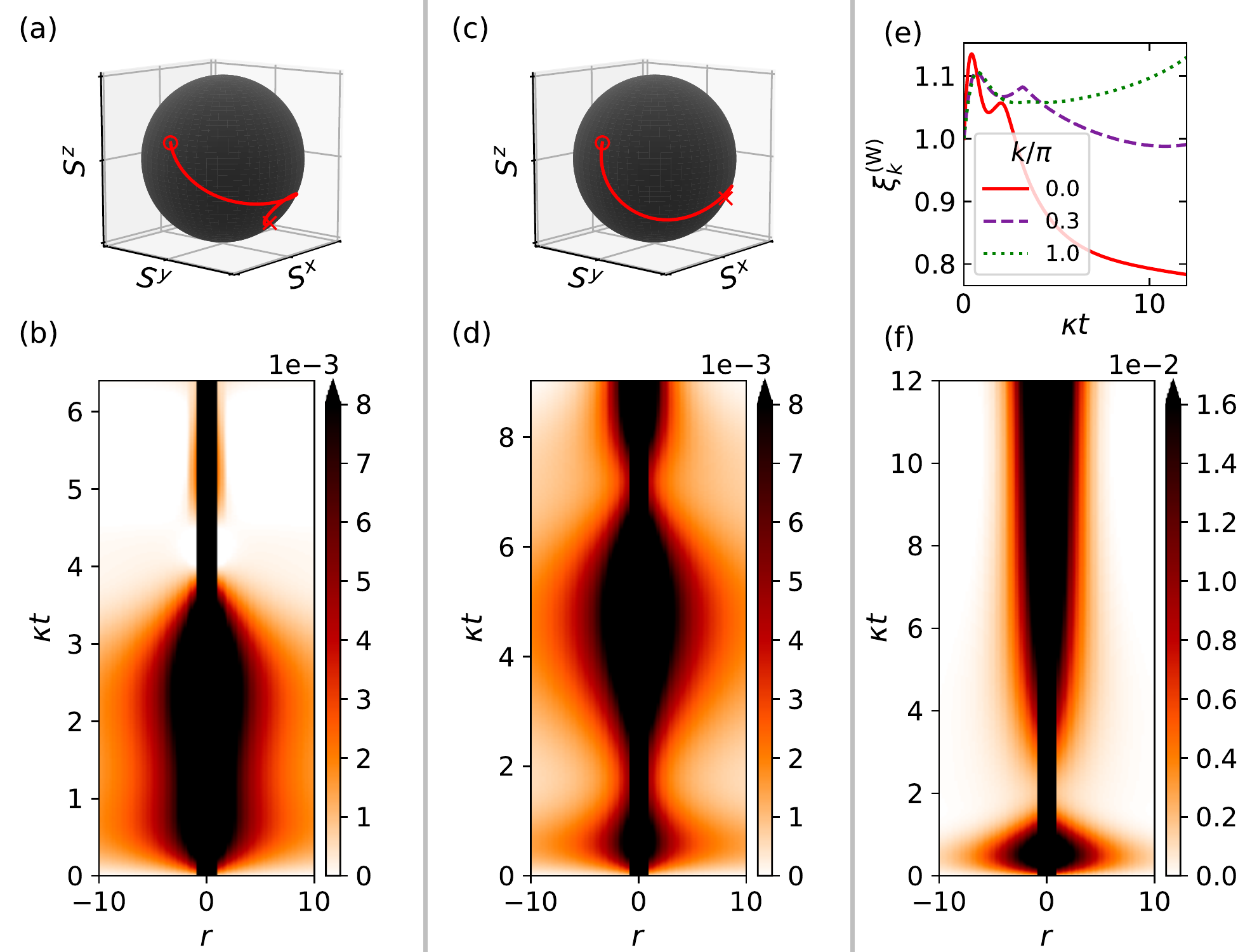}
	\caption{\textbf{Modulating correlations via a uniform field.} We initialize the system in a spin coherent state pointing in the direction $\theta(t=0)=0.4\pi$, $\phi(t=0)=0$ for all panels. \textbf{(a),(b)} Collective spin motion and connected correlation function $C^{zz}\left(r,t\right)$ for a long-range spatial profile with $\alpha=1.25$. System parameters are $\kappa=0.8$, $\omega_{F}=1.0$, and $\varphi=0.25\pi$. \textbf{(c),(d)} Collective spin motion and connected correlation function $C^{zz}\left(r,t\right)$ for a long-range spatial profile with $\alpha=1.1$. System parameters are $\kappa=0.95$, $\omega_{F}=1.0$, and $\varphi=0.1\pi$. \textbf{(e),(f)} Squeezing parameter and connected correlation function $C^{zz}\left(r,t\right)$ for a short-range spatial profile with $\chi=3.0$. System parameters are $\kappa=1.2$, $\omega_{F}=1.0$, and $\varphi=0$.}
	\label{fig:CorrPatterns}
\end{figure*}

 Figure~\ref{fig:ExpSchematic} schematically depicts how to realize non-local dissipation of spin-$1/2$ systems using cold atoms trapped in a single-mode optical cavity. The spin states are encoded in the hyperfine levels of the atoms and the cavity photon mode allows the atoms to communicate with each other, through both coherent interactions and non-local dissipation. There are three key components to this construction. First, a magnetic field gradient makes the energy of the hyperfine levels site-dependent and thereby endows the system with spatial resolution~\cite{PhysRevLett.123.130601,PhysRevX.8.011002,norcia2018cavity}. Second, a classical Raman beam with multiple sidebands provides control over atomic transitions between different spin states~\cite{hung2016quantum,PhysRevLett.123.130601}. The frequencies of the sidebands can be chosen so that communication between atoms via the cavity mode only depends on the distance between atoms, thereby enforcing translational invariance. The amplitudes of the sidebands determine the rate of internal atomic transitions and dictate the likelihood that two atoms a fixed distance apart communicate with each other, thereby setting the spatial profile of the dynamical channel. Third, cavity photon losses are large enough that the coherent spin-exchange contribution to dynamics is negligible and only dissipative dynamics remains. The three ingredients described above can be used to construct non-local dissipation channels with a variety of jump operators $\hat{L}_{n}$. Experiments will generally suffer from additional local dissipation arising from spontaneous scattering of individual atoms into free space; we derive conditions for the robustness of our set-up to such losses in the Supplemental Material (SM).
 
 In this work, we consider the case of a $\hat{L}_{n}=\hat{S}_{n}^{-}$ dissipation channel with $\hat{S}_{n}^{-}$ being the spin lowering operator on site $n$. Details of the experimental implementation for this channel and the construction a desired spatial profile $f_{n,m}$ are given in the SM. Here, we first examine the purely dissipative dynamics arising from Eq.~\ref{eq:model} with $\hat{H}=0$. We start the system in a coherent spin state in the northern hemisphere of the collective spin Bloch sphere, parameterized by the initial polar and azimuthal angles $\theta(t=0)=0.4\pi$ and $\phi(t=0)=0$, and compute the dynamics of the equal-time connected correlation function
\begin{equation}\label{eq:CorrFunc}
C^{zz}\left(r,t\right)=\langle \hat{S}_{n}^{z}\left(t\right)\hat{S}_{n+r}^{z}\left(t\right)\rangle -\langle \hat{S}_{n}^{z}\left(t\right)\rangle \langle \hat{S}_{n+r}^{z}\left(t\right)\rangle,
\end{equation}
which is sensitive to the action of spin losses generated by $\hat{L}_{n}=\hat{S}_{n}^{-}$ and directly measurable in experiment via state-selective fluorescence imaging~\cite{Periwal2021} .  We also track the evolution of the collective spin, $\langle\hat{\boldsymbol{S}}\rangle=\langle\sum_{n}\hat{\boldsymbol{S}}_{n}\rangle$, and the Kitagawa squeezing parameter, $\xi_{0}^{2}=\frac{4}{N}\text{\ensuremath{\min}}_{\boldsymbol{e}_{\perp}}\langle \Delta\left(\boldsymbol{e}_{\perp}\cdot\hat{\boldsymbol{S}}\right)^{2}\rangle $~\cite{kitagawa1993squeezed, ma2011quantum}, with minimization over directions $\boldsymbol{e}_{\perp}$ perpendicular to the mean spin direction $\boldsymbol{e}_{s}$. We compute the dynamics of these quantities for a thermodynamically large number of sites by extending the time-dependent self-consistent Hartree theory of spin waves developed in Refs.~~\cite{jamir, lerose2019prethermal, lerose2020origin} to the case of Lindblad channels~\cite{kushal}, and ensure that the total density of spin waves remains sufficiently small at every instant of time to ensure accuracy of the method. The dynamics for both long-range and short-range spatial profiles is shown in Fig.~\ref{fig:DispConfinement}. They share the same qualitative features; correlations spread for a period of time before contracting back towards an uncorrelated state as the collective spin crosses the equator of the Bloch sphere and eventually reaches the south pole. The squeezing parameter drops below one over the course of this motion thereby signifying entanglement in the system~\cite{sorensen2001many}. The main difference between long- and short-range profiles is that the short-range profile generates correlations that decay more quickly in space at any given time. 
These dynamics, and the time-dependent spin-wave theory used to compute them, are further characterized in Ref.~\cite{Seetharam2021dynamical}.

The spread and contraction of correlations is reminiscent of dynamical signatures of confinement in purely unitary spin systems~\cite{kormos2017real,PhysRevLett.122.150601, tan2019observation, PhysRevLett.124.180602}. There, correlations are confined due to bound states in the spectrum of Hamiltonianan which arise from an effective attractive potential for low-lying excitations. Here, however, confinement of correlations is an inherently non-equilibrium phenomenon stemming from the fact that non-local dissipation channels can both create and destroy correlations. 

Intuition for this property can be gained by examining the dynamics of the $\hat{L}_{n}=\hat{S}_{n}^{-}$ channel with long-range profile $f_{n,m}=\left(\left|n-m\right|+1\right)^{-\alpha}$, depicted in Fig.~\ref{fig:DispConfinement}(a)-(c). In the thermodynamic limit, the system behaves qualitatively as if $\alpha=0$ for any $\alpha\leq1$ and exhibits collective dynamics which is fully captured by the motion of the collective spin on a Bloch sphere~\cite{jamir}. For $\alpha$ just above $1$, the system is well-described by a collective spin moving along with spin-wave excitations generated on top of it by the spatially dependent spin-lowering jump operator. When the collective spin is in the northern hemisphere of the Bloch sphere, the average magnetization of the system is positive and the jump operator $\hat{S}_{n}^{-}$ creates spin-waves by lowering the magnetization away from that of a spin-coherent state which is fully polarized upwards. When the collective spin is in the southern hemisphere of the Bloch sphere, the average magnetization of the system is negative and the jump operator destroys spin-waves by lowering the average magnetization towards that of a spin-coherent state which is fully polarized downwards. The collective spin therefore acts as a mobile vacuum for excitations and its position controls whether the dissipation channel predominently creates or destroys correlations carried by these excitations. 





A uniform external field which guides the motion of the collective spin, and thereby influences when dissipation creates or destroys excitations,  can be used to modulate the spatio-temporal correlation pattern created by the $\hat{L}_{n}=\hat{S}_{n}^{-}$ dissipation channel. We demonstrate this control using a field of magnitude $\omega_{F}$ and direction $\varphi$ described by the Hamiltonian $\hat{H}=\omega_{F}\left(\cos\varphi\hat{S}^{x}+\sin\varphi\hat{S}^{z}\right)$, which generates the coherent part of dynamics in Eq.~\ref{eq:model}.


In Fig.~\ref{fig:CorrPatterns}(a)-(d), we show how the long-range confinement pattern of Fig~\ref{fig:DispConfinement}(a)-(c) can be modified. Figure~\ref{fig:CorrPatterns}(b) shows temporal control over the correlation pattern. The window of time during which the system remains correlated before decaying to an uncorrelated state is extended by a factor of approximately five.


Figure~\ref{fig:CorrPatterns}(d) shows that the confinement pattern can be modulated to exhibit oscillating correlations, which resembles a dissipation-induced limit cycle dressed by quantum fluctuations. This behaviour, however, is metastable and the system reaches a non-oscillatory steady state. 
For parameters $\alpha\leq1$, $\varphi=0$, and $\omega_F\geq\kappa$, the system is fully described by the classical motion of the collective spin and exhibits persistent oscillations~\cite{Fazio}. However, for $\alpha>1$, we find that these oscillations are eventually washed out by many-body fluctuations. 

In Fig.~\ref{fig:CorrPatterns}(e)-(f), we manipulate the short-range confinement pattern of Fig~\ref{fig:DispConfinement}(d)-(f) via the uniform field. Specifically, we send the system to an increasingly correlated state at late times in a fashion reminiscent of traditional dissipative state preparation schemes~\cite{diehl2008quantum}. 

 Our platform has potential utility for applications in quantum metrology and state preparation, which we explore by examining the
the finite wavevector squeezing parameter
\begin{equation}\label{eq:kSqueezingParam}
\left(\xi_{k}^{(\textrm{W})}\right)^{2} = \text{\ensuremath{\min}}_{\boldsymbol{e}_{\perp}}\biggr\{\frac{2^{\delta_{k\neq 0}}N\langle \Delta\left(\boldsymbol{e}_{\perp}\cdot\operatorname{Re}\{\hat{\boldsymbol{S}}_{k}\}\right)^{2}\rangle}{\lvert\langle \boldsymbol{e}_{s}\cdot\hat{\boldsymbol{S}}_\textrm{tot} \rangle\rvert^{2}}\biggr\}.
\end{equation}
where $\hat{\boldsymbol{S}}_{k}=\sum_{n}e^{-ikn}\hat{\boldsymbol{S}}_{n}$ and $\delta_{k\neq 0}=1$ if $k\neq0$ and $0$ otherwise. In the SM, we show that $\xi_{k}^{(\textrm{W})}$ is a generalization of the Wineland collective squeezing parameter~\cite{Wineland1992,ma2011quantum}, and quantifies metrologically useful entanglement when sensing a particular spatial mode of a spatially varying field.

Figure~\ref{fig:CorrPatterns}(e) shows that different wavevectors exhibit varying amounts of squeezing depending on the Fourier transform of the spatial profile, given by $\Gamma_{k}=\sum_{r=n-m} e^{ikr}f\left(\left|r\right|\right)$. Modes with larger $\Gamma_{k}$ will get squeezed more at short times. This fact can be exploited to preferentially squeeze a target mode $k^{*}$.

\begin{figure}[t!]
	\centering
	\includegraphics[width=0.99\columnwidth]{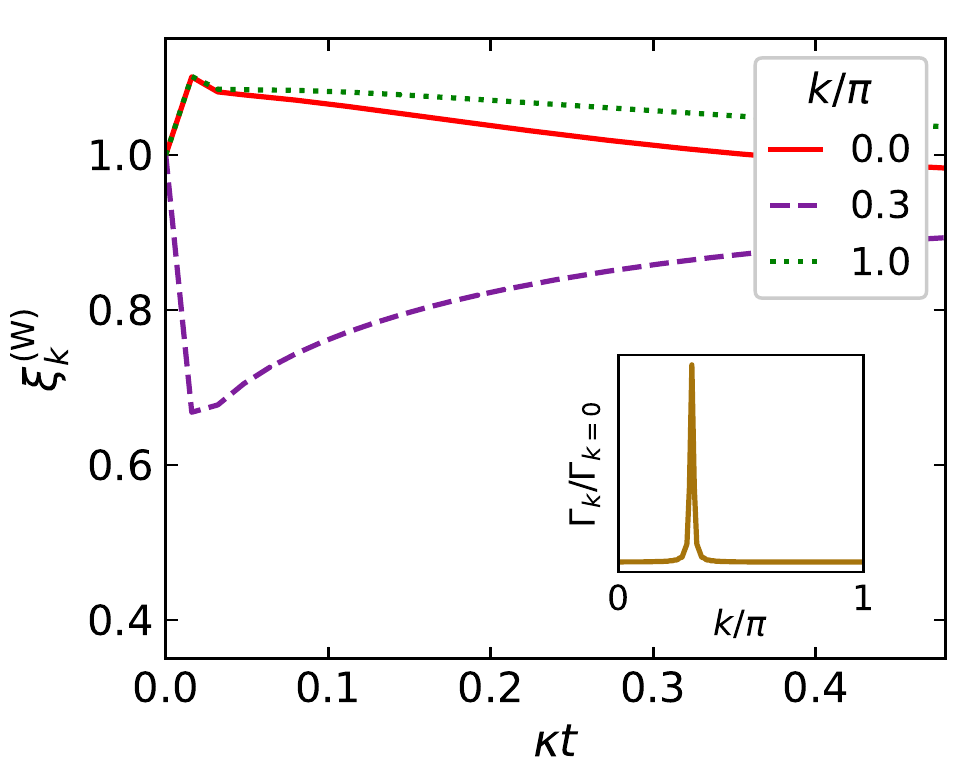}
	\caption{\textbf{Preferentially squeezing a target wavevector.} Spatial squeezing parameter for a spatial profile $f_{n,m}=\cos\left(k^{*}|n-m|\right)e^{-|n-m|/\chi}$ with $\chi=50$ and $k^{*}=0.3\pi$. System parameters are $\kappa=0.8$, $\omega_{F}=1.0$, and $\varphi=0$. The system is initialized in a spin coherent state pointing in the direction $\theta(t=0)=0.4\pi$, $\phi(t=0)=0$. The figure inset shows that the Fourier transform of the spatial profile is peaked around $k^{*}$.}
	\label{fig:kSqueezing}
\end{figure}

We demonstrate this control in Fig.~\ref{fig:kSqueezing}, where we show the dynamics of $\xi_{k}^{(\textrm{W})}$ for a spatial profile $f_{n,m}=\cos\left(k^{*}|n-m|\right)e^{-|n-m|/\chi}$ with $k^{*}=0.3\pi$. The figure inset shows that $\Gamma_{k}$ is a Lorentzian of width $\chi$ peaked at $k^{*}$. We see that the $k=k^{*}$ mode is squeezed more than other modes, including the collective $k=0$ mode which witnesses pairwise entanglement. Furthermore, squeezing at $k^{*}$ seems to antisqueeze other modes.




 Compared to purely unitary strategies, dissipative protocols may hold unique advantages in a variety of applications as irreversible dynamics is stable to variations in the initial state. We have checked, for instance, that the dynamical features shown in Fig.~\ref{fig:DispConfinement} and Fig.~\ref{fig:CorrPatterns} do not depend qualitatively on the choice of initial state. In contrast, the dynamical confinement of correlations in long-range interacting unitary quantum simulators requires careful preparation of the initial state~\cite{kormos2017real,PhysRevLett.122.150601, tan2019observation, PhysRevLett.124.180602}.
 
 The spread of correlations in  unitary long-range simulators has proven a fruitful area of inquiry for understanding entanglement dynamics in many-body systems~\cite{HaukeTagliacozzo,vodola2014kitaev,Gorshkov2}. By choosing a spatial profile $f_{n,m}=\left(\left|n-m\right|+1\right)^{-\alpha}$, our platform turns into a non-Hermitian analogue of such systems, thus opening an opportunity to explore how the purely dissipative character of dynamics affects entanglement spreading.



The correlation function in Eq.~\eqref{eq:CorrFunc} can be experimentally measured by state-selective fluorescence imaging~\cite{Periwal2021}. Initially preparing a spin texture alternatively allows one to track the dynamics of the system with direct measurements of the local magnetization~\cite{PhysRevLett.125.060402}, potentially revealing novel transport mechanisms assisted by non-local dissipation. Our platform also offers the prospect of studying quantum information scrambling~\cite{marino2019cavity, PhysRevX.9.041011} and novel phase transitions~\cite{dogra2019dissipation} in purely dissipative cavity QED simulators.

Furthermore, the ability to squeeze the system at desired wavevectors may be useful for spatially-resolved 
magnetometry, thus providing an advantage over systems employing homogeneous, collective dissipation~\cite{PhysRevLett.110.120402}, which can only squeeze the collective spin mode. Realizing our platform's potential for spatial magnetometry requires optimizing the choice of profile, $f_{n,m}$, and jump operator, $\hat{L}_{n}$, characterizing the non-local dissipation channel to maximally decrease the value of $\xi_{k}^{(\textrm{W})}$ within the Heisenberg limit (see the SM). Feedback conditioned on emitted photons and use of ensembles of atoms may offer additional routes towards increased metrological sensitivty~\cite{PhysRevLett.125.060402, PhysRevA.81.021804,PhysRevLett.104.073602, cox2016deterministic,kroeger2020continuous}. \\





We thank G. Bentsen, P. Calabrese, E. Davis, E. Demler, M. Schleier-Smith, and S. Zeytino\"u{g}lu for stimulating discussions. K.S. conducted this research with Government support under and awarded by DoD, Air Force Office of Scientific Research, National Defense Science and Engineering Graduate (NDSEG) Fellowship, 32 CFR 168a. J.M. was supported by the European Union’s Horizon 2020 research and innovation programme under the Marie Sklodowska-Curie grant agreement No 745608 (QUAKE4PRELIMAT), and by  the Deutsche Forschungsgemeinschaft (DFG, German Research Foundation) -- Project-ID 429529648 -- TRR 306 QuCoLiMa ("Quantum Cooperativity of Light and Matter''). A.L. acknowledges funding by the Swiss National Science Foundation. R.F. acknowledges partial financial support from the Google Quantum Research Award. R.F.'s work has been conducted within the framework of the Trieste Institute for Theoretical Quantum Technologies (TQT). 


%

\pagebreak
\onecolumngrid

\renewcommand{\theequation}{S.\arabic{equation}}
\renewcommand{\thesection}{S\arabic{section}}
\renewcommand{\thefigure}{S\arabic{figure}}

\setcounter{equation}{0}
\setcounter{figure}{0}

\section*{\large Supplementary Information}


In this Supplementary Material (SM), we give an experimental construction of a non-local dissipator corresponding to a $\hat{L}_{n}=\hat{S}_{n}^{-}$ loss channel with a translationally invariant spatial profile $f(\abs{n-m})$, as described in Eq. (1) of the main text.  Our proposal is motivated by experiments employing clouds of  $^{87}$Rb atoms coupled to a single photon mode in an optical cavity. In previous works, the cavity mode can be employed as a resource to mediate spin-exchange coherent interactions among the atoms~\cite{Monika}, which can be accompanied by collective dissipation depending on the cooperativity of the cavity. Here we work in a complementary limit and engineer incoherent spin emission with spatial resolution. The premise of our construction is to take a chain of atoms, each with three hyperfine levels out of which two are degenerate, trap them inside an optical cavity, and then apply a magnetic field gradient and a Raman beam to the system with several sidebands of tunable frequency and amplitude (see  Ref.~\cite{Cirac} for a related implementation in photonic waveguides). The magnetic field gradient splits the degeneracy of each atom such that its energy levels form a $\Lambda$-configuration; the energies are site-dependent and make the atoms spatially distinct. The Raman beam couples one leg of the $\Lambda$-configuration, while the cavity mode couples the other. The cavity mode mediates communication between atoms at different sites, allowing for both coherent atom-atom interactions as well as  indistinguishable atomic losses. The choice of frequencies and amplitudes of the sidebands comprising the Raman beam dictates the probability that atoms at different sites communicate with each other through the cavity photon, thereby setting the spatial profile which shapes both coherent interactions and losses. If the cavity is made to be sufficiently leaky, the coherent interactions are washed out and only dissipative dynamics with the desired spatial profile remains. 

In the final section of the SM, we show how the finite wavevector squeezing parameter described in the main text quantifies metrologically useful entanglement when sensing a specific Fourier component of a spatially varying field.


\section{Non-local losses}

We now give a detailed construction of the experimental implementation. We consider a one-dimensional chain of $N$ atoms labeled by lattice index $n=1, .., N$. Each atom has two internal states
$\ket{\tilde{g}}_{n}$ and $\ket{e}_{n}$. The state $\ket{\tilde{g}}_{n}$
belongs to a degenerate hyperfine manifold which, under application of an external field, splits as $\ket{\tilde{g}}_{n}\rightarrow\left\{ \ket{s}_{n},\ket{g}_{n}\right\}$. We encode the spin $1/2$ Hilbert space $\{\ket{\uparrow}_{n}, \ket{\downarrow}_{n}\}$ in this ground state manifold.
We take $\ket{s}_{n}$ to be the lower energy state and set its energy
to zero without loss of generality. The energy difference between
$\ket{s}_{n}$ and $\ket{g}_{n}$ is given as $\omega_{g,n}$ and we will refer to 
the energy difference between $\ket{s}_{n}$ and $\ket{e}_{n}$ 
as $\omega_{e,n}$. These energies are position dependent since  they inherit spatial dependence from the external applied magnetic field gradient. In terms of the operators $\hat{\sigma}_{ab}^{n}=\ket{a}_{n}\bra{b}_n$,
with $a,b\in\left\{ s,g,e\right\} $, the bare atomic Hamiltonian
reads
\begin{equation}
\hat{H}_{\text{a}}=\sum_{n}\left(\omega_{e,n}\hat{\sigma}_{ee}^{n}+\omega_{g,n}\hat{\sigma}_{gg}^{n}\right).
\end{equation}
We now dipole-couple the states $\ket{s}_{n}$ and $\ket{e}_{n}$
using a Raman driving field $\tilde{\Omega}\left(t\right)=\sum_{\alpha=0}^{m_{p}-1}\Omega_{\alpha}e^{i\omega_{\alpha}t}$
where $\omega_{\alpha}$ represents each of the $m_{p}$ different
drive frequencies and $\Omega_{\alpha}$ represents the Rabi frequency
(beam amplitude) associated with those frequencies. We can define the
frequency $\omega_{L}\equiv\omega_{\alpha=0}$ as the main frequency
and rewrite the driving field in terms of the detunings $\tilde{\omega}_{\alpha}\equiv\omega_{\alpha}-\omega_{L}$
as $\tilde{\Omega}\left(t\right)=\Omega\left(t\right)e^{i\omega_{L}t}$
where $\Omega\left(t\right)\equiv\sum_{\alpha=0}^{m_{p}-1}\Omega_{\alpha}e^{i\tilde{\omega}_{\alpha}t}$.
Note that $\tilde{\omega}_{\alpha}=0$ by definition. The dipole coupling
between $\ket{s}_{n}$ and $\ket{e}_{n}$ is then described by the Hamiltonian
\begin{equation}
\hat{H}_{\text{d}}=\sum_{n}\left(\frac{\Omega\left(t\right)}{2}e^{i\omega_{L}t}\hat{\sigma}_{se}^{n}+\frac{\Omega^{*}\left(t\right)}{2}e^{-i\omega_{L}t}\hat{\sigma}_{es}^{n}\right).
\end{equation}
We now consider an optical cavity mode that dipole couples $\ket{g}_{n}$
and $\ket{e}_{n}$. The photon mode, represented by the operator
$\hat{a} $, has a frequency $\omega_{c}$ and couples to atom $n$ through the single-photon coupling $g$. The bare photon Hamiltonian and the light-matter coupling between atoms and photons are  given respectively by
\begin{align}
\hat{H}_{\text{p}} & =\omega_{c}\hat{a}^{\dagger}\hat{a},\\
\hat{H}_{\mathbf{\text{lm}}} & =\sum_{n}\left(g\hat{a}\hat{\sigma}_{eg}^{n}+g^{*}\hat{a}^{\dagger}\hat{\sigma}_{ge}^{n}\right).
\end{align}
The total density matrix of the system has dynamics given by the quantum master equation in Lindblad form
\begin{equation}\label{eq:Lindblad}
\frac{d}{dt}\rho=-i\left[\hat{H},\rho\right]+\mathcal{D}_{\text{leak}}\left(\rho\right).
\end{equation}
The last term is the dissipator corresponding to photon losses  occurring with rate $\gamma$
\begin{equation}\label{eq:Dissipator}
\mathcal{D}_{\text{leak}}\left(\rho\right)  =\gamma\left(\hat{a}^{\dagger}\rho\hat{a}-\frac{1}{2}\left\{ \hat{a}\hat{a}^{\dagger},\rho\right\} \right).
\end{equation}
When the excited state $|e\rangle$ is largely detuned by $\Delta=|\omega_e - \omega_L|$ from the other atomic and photonic energy scales ($\Delta\gg\Omega_{\alpha},g$), one can use a Schrieffer-Wolf transformation to eliminate the state and write an effective Hamiltonian for the remaining atomic Hilbert space.  The light-matter interaction coupling at leading order in $1/\Delta$ then becomes

\begin{equation}\label{interact}
\hat{H}_{\text{lm}}  =-\sum_{n}\left(\frac{g\Omega\left(t\right)}{\Delta}\hat{a}\hat{\sigma}_{sg}^{n}e^{i\left(\omega_{L}-\omega_{c}-\omega_{g,n}\right)t}+\frac{g^{*}\Omega^{*}\left(t\right)}{\Delta}\hat{a}^{\dagger}\hat{\sigma}_{gs}^{n}e^{-i\left(\omega_{L}-\omega_{c}-\omega_{g,n}\right)t}\right).
\end{equation}
Defining, $\eta_{\alpha} \equiv \frac{\Omega_{\alpha} g}{\Delta}$ and $\delta_{\alpha,n} \equiv\omega_{\alpha} - \omega_{g,n} - \omega_c$, the interaction Hamiltonian can be written as
\begin{align}\label{eq:LMInteraction}
\hat{H}_{\text{lm}}=-\sum_{n,\alpha}\left[\eta_{\alpha} \hat{a}\hat{\sigma}_{sg}^{n}e^{i\delta_{\alpha,n}t}+ h.c.\right].
\end{align}

The cavity mode mediates communication between atoms. We now assume that the cavity photon loss is large enough that
\begin{equation}\label{eq:adiabaticElimConds}
\begin{split}
\gamma\gg \eta_{\alpha}, \quad
\gamma\gg \delta_{\alpha,n},
\end{split}
\end{equation}
and therefore the cavity photon loss occurs on a timescale much faster than the effective dynamics of the spins. The light field can then be adiabatically eliminated and becomes enslaved to atomic operators~\cite{Agarwal}. The Heisenberg evolution of the light field can then be expanded in powers of $\epsilon$, with $\epsilon=\eta_{\alpha}/\gamma$ or $\epsilon=\delta_{\alpha,n}/\gamma$: 
\begin{equation}\label{slave}
\begin{split}
\hat{a}\left(t\right) = i2\sum_{n,\alpha}\left(\frac{\eta^*_{\alpha}}{\gamma} \hat{\sigma}_{gs}^{n}e^{-i\delta_{\alpha,n}t}\right)
+2\sum_{n,\alpha}\left(\frac{\delta_{\alpha,n}\eta^*_{\alpha}}{\gamma^{2}} \hat{\sigma}_{gs}^{n}e^{-i\delta_{\alpha,n}t}\right)+\mathcal{O}\left(\epsilon^3\right),
\end{split}
\end{equation}

Before using the above expression to replace the light field in the full Lindblad equation, Eq.~\eqref{eq:Lindblad}, we can gain insight into the effective dynamics of the system after elimination of the light field by performing this substitution for the equation of motion of a single spin operator:
\begin{align}\label{eq:spinEoM}
\frac{d}{dt}\hat{\sigma}_{gs}^n &= i\hat{a}\;\hat{\sigma}_{z}^n \sum_{\alpha}\eta_{\alpha}e^{i\delta_{\alpha,n}t}\nonumber\\
&\rightarrow -\hat{\sigma}_{z}^n \sum_{m,\alpha,\beta}  \left(\gamma_{\mathrm{eff}}\right)_{\alpha,\beta} \left(1-i\frac{\delta_{\beta,m}}{\gamma}\right) e^{i(\delta_{\alpha,n}-\delta_{\beta, m})t}  \hat{\sigma}_{gs}^m + \mathcal{O}\left(\epsilon^3\right),
\end{align}
where we have defined $\left(\gamma_{\mathrm{eff}}\right)_{\alpha,\beta}  \equiv  2{\eta^*_{\alpha}\eta_{\beta} }/{\gamma}$. We see that the motion of the $n^{\mathrm{th}}$ atom is conditioned by  the motion of the $m^{\mathrm{th}}$ one, with $\left(\gamma_{\mathrm{eff}}\right)_{\alpha,\beta}$ setting the effective coupling rate. The leading order contribution to the motion is dissipative dynamics with rate $\left(\gamma_{\mathrm{eff}}\right)_{\alpha,\beta}$, with the subleading contribution being coherent dynamics with frequency $\left(\gamma_{\mathrm{eff}}\right)_{\alpha,\beta}\frac{\delta_{\beta,m}}{\gamma}$. When the effective coupling constant, $\left(\gamma_{\mathrm{eff}}\right)_{\alpha,\beta}$, is much smaller than the 
minimum detuning between the atomic transition frequencies, we can ignore the off-resonant couplings and only consider the interaction 
between atoms $n$ and $m$ for which $\delta_{\alpha,n}-\delta_{\beta, m} = 0$. Specifically, we require
\begin{equation}\label{eq:fourPhotonRes}
\left(\gamma_{\mathrm{eff}}\right)_{\alpha,\beta}\ll \textrm{min}\{\delta_{\alpha,n}-\delta_{\beta, m}\},
\end{equation}
where the minimization means the smallest nonzero value of $\delta_{\alpha,n}-\delta_{\beta, m}$. Formally, Eq.~\eqref{eq:fourPhotonRes} is derived by taking the long time average of~\eqref{eq:spinEoM}, and then applying the Sokhotski–Plemelj lemma to extract the singular part of the time integral (resonant process) and the regular part (off-resonant processes). The contribution of the off-resonant term becomes negligible  when the condition~\eqref{eq:fourPhotonRes} is satisfied (see Ref.~\cite{Cirac}). We can then safely restrict the dynamics to the resonance shell $\delta_{\alpha, n}=\delta_{\beta, m}$, which can be restated as
\begin{equation}\label{reson}
\omega_{g,m}-\omega_{g,n}=\tilde{\omega}_\beta-\tilde{\omega}_\alpha.
\end{equation} 
In order to introduce spatial addressability in the system, we choose the site-dependent energy shifts  as $\omega_{g,n}=\mu n$, which is implemented via an externally imposed linear 
magnetic field. 
We also choose the sideband detunings as $\tilde{\omega}_{\alpha}=\mu\alpha$. After our choice of sideband detunings, the resonance condition Eq.~\eqref{reson} reads 
\begin{equation}\left(\alpha-\beta\right)=\left(n-m\right).\end{equation} 
This selection rule makes pairs of atoms at distance $n-m$ apart interact. The dynamics of a single spin, given by Eq.~\ref{eq:spinEoM}, then becomes
\begin{equation}\label{eq:spinEoM_res}
\frac{d}{dt}\hat{\sigma}_{gs}^n \approx \rightarrow -\hat{\sigma}_{z}^n \sum_{m,\beta}  \left(\gamma_{\mathrm{eff}}\right)_{\beta+(n-m),\beta} \left(1-i\frac{\delta_{\beta,m}}{\gamma}\right) \hat{\sigma}_{gs}^m,
\end{equation}
and we see that the effective coupling rate, $\left(\gamma_{\mathrm{eff}}\right)_{\beta+(n-m),\beta}$, depends only on the distance between atoms $n$ and $m$. The leading order term, corresponding to dissipative dynamics, is therefore translationally invariant. The subleading coherent term, proportional to $\delta_{\beta,m}/\gamma$, however, does have explicit position dependence. We note that if the condition in Eq.~\eqref{eq:fourPhotonRes} is violated, then atoms on multiple sites can communicate and even the dissipative dynamics will not be translationally invariant. 

We now perform this same adiabatic elimination of the cavity photon on the full Lindblad equation, Eq.~\eqref{eq:Lindblad}, by replacing $\hat{a}$ with the expression in Eq.~\eqref{slave}. Keeping terms up to $\mathcal{O}(\epsilon^2)$ in Eq.~\eqref{slave}, the dissipator given by Eq.~\eqref{eq:Dissipator} becomes
\begin{equation}
\mathcal{D}_{\text{leak}}\left(\rho\right)\approx 
2\sum_{n,m,\alpha,\beta}\left(\gamma_{\mathrm{eff}}\right)_{\alpha,\beta}\left(1+i\frac{(\delta_{\alpha,n}-\delta_{\beta, m})}{\gamma}\right)e^{i(\delta_{\alpha,n}-\delta_{\beta, m})t}  \left(\hat{\sigma}_{gs}^{m}\rho\hat{\sigma}_{sg}^{n}-\frac{1}{2}\left\{ \hat{\sigma}_{sg}^{n}\hat{\sigma}_{gs}^{m},\rho\right\} \right),
\end{equation}
while the coherent light-matter interaction, given by Eq.~\eqref{eq:LMInteraction}, becomes
\begin{equation}
\hat{H}_{\text{lm}}\approx-i2\sum_{n,m,\alpha,\beta}\left(\gamma_{\mathrm{eff}}\right)_{\alpha,\beta}e^{i(\delta_{\alpha,n}-\delta_{\beta, m})t}\left(\hat{\sigma}_{sg}^{n}\hat{\sigma}_{gs}^{m}-\hat{\sigma}_{sg}^{n}\hat{\sigma}_{gs}^{m}\right)
-2\sum_{n,m,\alpha,\beta}\left(\gamma_{\mathrm{eff}}\right)_{\alpha,\beta}\frac{\delta_{\beta,m}}{\gamma}e^{i(\delta_{\alpha,n}-\delta_{\beta, m})t}\hat{\sigma}_{sg}^{n}\hat{\sigma}_{gs}^{m}.
\end{equation}
We see that the first term in the above equation vanishes and we are left with
\begin{equation}
\hat{H}_{\text{lm}}\approx-2\sum_{n,m,\alpha,\beta}\left(\gamma_{\mathrm{eff}}\right)_{\alpha,\beta}\frac{\delta_{\beta,m}}{\gamma}e^{i(\delta_{\alpha,n}-\delta_{\beta, m})t}\hat{\sigma}_{sg}^{n}\hat{\sigma}_{gs}^{m}.
\end{equation}
The master equation for the density matrix describing the system can thus be written as
\begin{align}
\frac{d}{dt}\rho  &\approx 
2\sum_{n,m,\alpha,\beta}\left(\gamma_{\mathrm{eff}}\right)_{\alpha,\beta}\left(1+i\frac{(\delta_{\alpha,n}-\delta_{\beta, m})}{\gamma}\right)e^{i(\delta_{\alpha,n}-\delta_{\beta, m})t}  \left(\hat{\sigma}_{gs}^{m}\rho\hat{\sigma}_{sg}^{n}-\frac{1}{2}\left\{ \hat{\sigma}_{sg}^{n}\hat{\sigma}_{gs}^{m},\rho\right\} \right)+\label{eq:DispPreRes}\\
&+i2\sum_{n,m,\alpha,\beta}\left(\gamma_{\mathrm{eff}}\right)_{\alpha,\beta}\frac{\delta_{\beta,m}}{\gamma}e^{i(\delta_{\alpha,n}-\delta_{\beta, m})t}\left[\hat{\sigma}_{sg}^{n}\hat{\sigma}_{gs}^{m},\rho\right].    
\end{align}
If the condition Eq.~\eqref{eq:fourPhotonRes} is satisfied, then we can restrict dynamics to the resonance shell defined by Eq.~\eqref{reson} and the resulting Lindblad equation is
\begin{equation}\label{eqrho}
\begin{split}
\frac{d}{dt}\rho  &\approx 
\left(\frac{4|g|^2 \Omega^2_M}{\gamma\Delta^{2}}\right)\sum_{n,m}f_{n,m}\left(\hat{\sigma}_{gs}^{m}\rho\hat{\sigma}_{sg}^{n}-\frac{1}{2}\left\{ \hat{\sigma}_{sg}^{n}\hat{\sigma}_{gs}^{m},\rho\right\} \right)
+i\left(\frac{4|g|^2\Omega^2_M}{\gamma\Delta^{2}}\right)\sum_{n,m}\left[\tilde{f}_{n,m}\hat{\sigma}_{sg}^{n}\hat{\sigma}_{gs}^{m},\rho\right]
\end{split}
\end{equation}
where 
\begin{equation}\label{couple}
\begin{split}
f_{n,m}  =\sum_{\left(\alpha,\beta\right);\left(\alpha-\beta\right)=\left(n-m\right)}\frac{\Omega_{\alpha}\Omega_{\beta}^{*}}{\Omega^2_M}, 
\quad{\tilde{f}}_{n,m}   =\sum_{\left(\alpha,\beta\right);\left(\alpha-\beta\right)=\left(n-m\right)}\frac{\delta_{\beta,m}}{\gamma}\frac{\Omega_{\alpha}\Omega_{\beta}^{*}}{\Omega^2_M}.
\end{split}
\end{equation}
In the above equation, we define $\Omega_M$ as the largest Raman sideband amplitude and choose it as a representative scale for the drive. Note that the term in Eq.~\eqref{eq:DispPreRes} that is proportional to $\delta_{\alpha,n}$ exactly cancels the term proportional to $\delta_{\beta,m}$ when we satisfy the two-atom resonance condition Eq.~\eqref{reson}.

Collating the conditions, Eq.~\eqref{eq:adiabaticElimConds} and Eq.~\eqref{eq:fourPhotonRes}, for adiabatically eliminating the cavity photon, we have:
\begin{equation}\label{eq:treatmentConditions}
\eta_{\alpha} \ll \gamma,
\quad
\delta_{\alpha,n} \ll \gamma,
\quad
\frac{\eta_{\alpha}\eta_{\beta}^{*}}{\textrm{min}\{\delta_{\alpha,n}-\delta_{\beta, m}\}} \ll \gamma.
\end{equation}
Recall that in the third condition above, the denominator, $\textrm{min}\{\delta_{\alpha,n}-\delta_{\beta, m}\}$, is a minimization over processes with $\delta_{\alpha,n}\neq\delta_{\beta,m}$. For any fixed $\eta_{\alpha}$ and $\delta_{\alpha,n}$, a sufficiently large cavity decay $\gamma$ allows all three conditions to be simultaneously satisfied. Later, we show that these conditions can be consistently satisfied by providing numerical estimates using parameters from cavity QED experiments. We note that the condition $\delta_{\alpha,n} \ll \gamma$ is satisfied by choosing a large cavity decay, $\gamma$, and a non-zero $\delta_{\alpha,n}$. In fact, we require that $\delta_{\alpha,n}\neq0$ in order to preserve spatial structure in the dynamics. We can see this by setting $\delta_{\alpha,n}=0$ in Eq.~\eqref{slave} to get $\hat{a}\left(t\right) = i \zeta \hat{\sigma}_{gs}$ where $\hat{\sigma}_{gs}=\sum_{n}\hat{\sigma}_{gs}^{n}$ and $\zeta=\frac{2}{\gamma}\sum_{\alpha}\eta^*_{\alpha}$. Single-atom resonant processes, characterized by $\delta_{\alpha,n}=0$, therefore only lead to collective dissipation of all spins in the system, arising from collective emission of the cavity photon, rather than non-local dissipation with spatial structure.

Physically, we can interpret the set-up resulting in Eq.~\eqref{eqrho} as follows. Both non-local dissipation and coherent interactions amongst the spins are mediated by non-resonant virtual photons, corresponding to $\delta_{\alpha,n}\neq0$, which satisfy the two-atom resonance condition $\delta_{\alpha,n}=\delta_{\beta,m}$.  The spatial profile of the resulting non-local dissipation, $f_{n,m}$ , is translationally invariant and represents leakage of a cavity photon without certainty about which of the two atoms, $n$ or $m$, it came from. The spatial profile of the coherent interaction, $\tilde{f}_{n,m}$, represents a spin-exchange between the atoms which is suppressed by a factor $\delta_{\beta,m}/\gamma$ due to the highly lossy cavity. The conditions in Eq.~\eqref{eq:treatmentConditions} represent a regime where the cavity loss, $\gamma$, is large enough that: (i) the effective dynamics of each individual spin, occurring through a $\Lambda$-process in the atom with rate $\eta_{\alpha}$, occurs slowly compared to the cavity photon loss so the photon only serves to mediate coherent interactions and non-local emission from pairs of spins, (ii) coherent interactions of the spins are suppressed, and (iii) the time-scale of the effective non-local emission from pairs of spins, set by  $\tau=1/\left(\gamma_{\mathrm{eff}}\right)_{\alpha,\beta}  \propto  \gamma/\eta^*_{\alpha}\eta_{\beta}$, is slow enough that off-resonant two-atom processes ($\delta_{\alpha,n}\neq\delta_{\beta,m}$) average to zero and only the resonant two-atom process remains. This resonant two-atom process is a translationally-invariant non-local emission from pairs of atoms.

The quantity $f_{n,m}=f\left(\abs{n-m}\right)$ only depends on the difference $n-m$ and sets the spatial profile of the non-local dissipation. We can design the desired translationally-invariant profile
$f_{n,m}=f\left(\abs{n-m}\right)$ of the dissipator by exactly solving $f\left(\abs{n-m}\right)=\sum_{\left(\alpha,\beta\right);\left(\alpha-\beta\right)=\left(n-m\right)}\Omega_{\alpha}\Omega_{\beta}^{*}/\Omega^2_M$. Later, we explicitly show how to numerically invert this equation. Defining $\kappa={\left|g\right|^{2}\Omega^2_M}/({\gamma\Delta^{2}})$ and relabeling the projection operators as $\hat{S}_{n}^{-}=  \hat{\sigma}_{gs}^{n}/2$ and
$\hat{S}_{n}^{+}= \hat{\sigma}_{sg}^{n}/2$, we have
\begin{equation}\label{eq:effectiveModel}
\begin{split}
\frac{d}{dt}\rho  &\approx 
\kappa\sum_{n,m}f\left(\left|n-m\right|\right)\left(\hat{S}_{m}^{-}\rho\hat{S}_{n}^{+}-\frac{1}{2}\left\{ \hat{S}_{n}^{+}\hat{S}_{m}^{-},\rho\right\} \right)
+i\left[\kappa\sum_{n,m}\tilde{f}_{n,m}\hat{S}_{n}^{-}\hat{S}_{m}^{+},\rho\right]
\end{split}
\end{equation}
The first term in Eq.~\eqref{eq:effectiveModel} is the desired non-local dissipation while the second term represents coherent spin-exchange interactions mediated by a virtual photon emitted in the  $\Lambda$-process within one atom and absorbed via the reverse $\Lambda$-process in a second atom a distance $|n-m|$ away~\cite{Cirac,treelike}. Comparing the expressions for $\tilde{f}_{n,m}$ and $f_{n,m}$ in Eq.~\eqref{couple}, we see that the coherent dynamics are subleading to the dissipative dynamics. Therefore, when the cavity decay is large enough that $\gamma\gg\delta_{\beta,m}$, the coherent dynamics vanishes and we are left with purely dissipative dynamics with a spatial profile $f\left(\abs{n-m}\right)$:
\begin{equation}\label{eq:dissipativeModel}
\begin{split}
\frac{d}{dt}\rho  &=
\kappa\sum_{n,m}f\left(\left|n-m\right|\right)\left(\hat{S}_{n}^{-}\rho\hat{S}_{m}^{+}-\frac{1}{2}\left\{ \hat{S}_{m}^{+}\hat{S}_{n}^{-},\rho\right\} \right)
\end{split}
\end{equation}
which is the non-local $\hat{L}_{n}=\hat{S}_{n}^{-}$ loss channel we aimed to construct. The positivity of this Lindblad map is guaranteed by the positivity of the Raman sideband amplitudes $\Omega_{\beta}$ that determine $f\left(\left|n-m\right|\right)$; we require $f\left(\left|n-m\right|=0\right)\neq0$ to ensure a positive Lindblad map, which is violated only when all sideband amplitudes are zero and we have no dynamics.

\begin{figure*}[t!]
	\centering
	\includegraphics[width=0.99\textwidth]{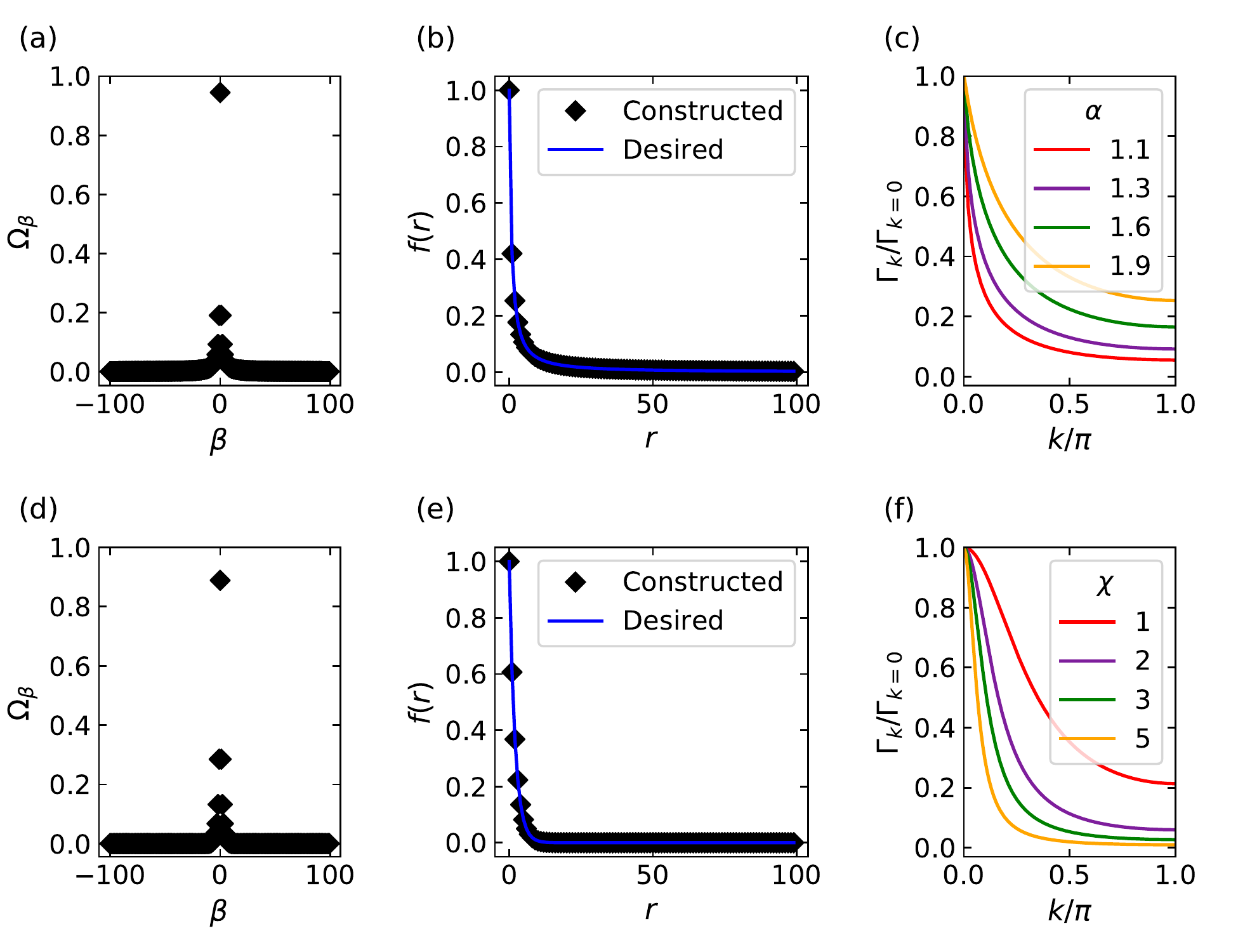}
	\caption{\textbf{Sideband amplitude construction.} \textbf{(a)} Sideband amplitudes required to construct a long-range spatial profile $f\left(\abs{n-m}\right)=\left(\left|n-m\right|+1\right)^{-\alpha}$ on a 100 site chain. \textbf{(b)} Spatial profile resulting from the amplitudes in (a). \textbf{(c)} Fourier transform of the long-range spatial profile. (d) Sideband amplitudes required to construct a short-range spatial profile $f\left(\abs{n-m}\right)=e^{-|n-m|/\chi}$ on a 100 site chain. \textbf{(e)} Spatial profile resulting from the amplitudes in (d). \textbf{(f)} Fourier transform of the short-range spatial profile.}
	\label{fig:Exp_SidebandConstruction}
\end{figure*}

\begin{figure*}[t!]
	\centering
	\includegraphics[width=0.99\textwidth]{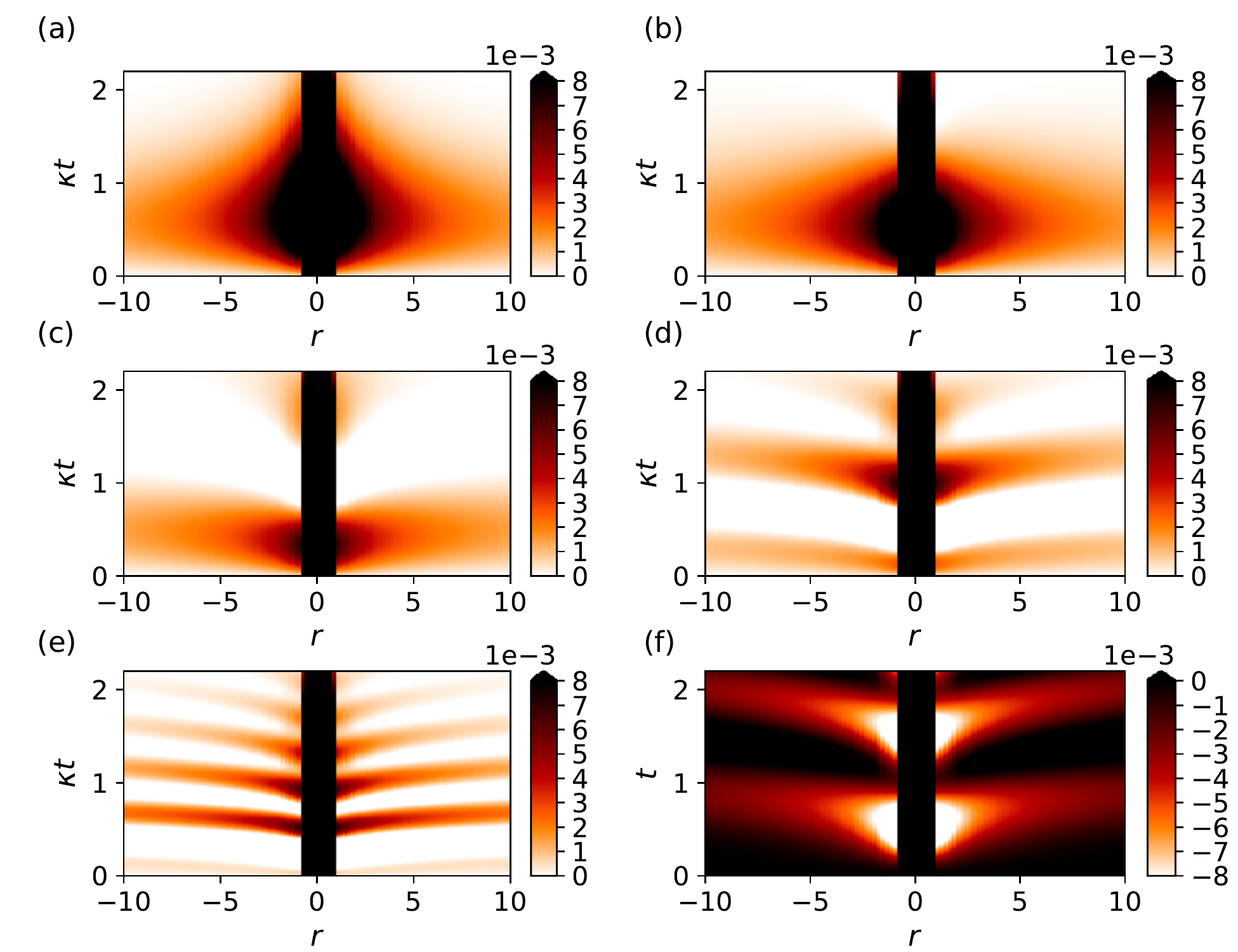}
	\caption{\textbf{Crossover between dissipative and coherent dynamics.} We compute the correlation function $C^{zz}\left(r,t\right)$ for a system whose dynamics is comprised of a long-range dissipation channel ($\hat{L}_{n}=\hat{S}_{n}^{-}$) with strength $\kappa$ and a long-range spin-exchange Hamiltonian with strength $\eta$. For both generators, the spatial profile is $f\left(\abs{n-m}\right)=\left(\left|n-m\right|+1\right)^{-\alpha}$ with $\alpha=1.25$. The system exhibits a crossover between $\eta/\kappa=0.25$ and $\eta/\kappa=0.5$. \textbf{(a)} $\kappa=1.0$, $\eta=0.0$ (only dissipation). \textbf{b} $\kappa=1.0$, $\eta=0.25$. \textbf{c} $\kappa=1.0$, $\eta=0.5$. \textbf{d} $\kappa=1.0$, $\eta=1.0$. \textbf{e} $\kappa=1.0$, $\eta=2.0$. \textbf{f} $\kappa=0.0$, $\eta=1.0$ (only Hamiltonian).}
	\label{fig:Exp_LRHam}
\end{figure*}

\section{Engineering the spatial profile}

We now show how to construct a desired dissipation profile $f\left(\abs{n-m}\right)$
by choosing the Raman sideband amplitudes $\left\{ \Omega_{\beta}\right\} $
appropriately. In this section, we work in units where the maximum sideband amplitude is normalized to $1$ ($\Omega_M=1$). The
equation we want to invert is
\begin{align*}
f\left(\abs{n-m}\right) & =\sum_{\alpha,\beta;\left(\alpha-\beta\right)=\left(n-m\right)}\Omega_{\alpha}\Omega_{\beta}^{*}\\
& =\sum_{\beta}\Omega_{\beta+\left(n-m\right)}\Omega_{\beta}^{*},
\end{align*}
with $r\equiv n-m$. For a system of $N$ spins, we have $r=-\left(N-1\right),-\left(N-2\right),..,0,...,\left(N-2\right),\left(N-1\right)$.
Recalling that $f\left(\abs{n-m}\right)=f\left(r\right)$ is a translationally invariant
profile, we have
\begin{equation}\label{eq:autoCorr}
f\left(r\right)=\sum_{\beta}\Omega_{\beta+r}\Omega_{\beta}^{*},
\end{equation}
where we will need $2N-1$ sidebands corresponding to the 
values that $r$ can take; the sidebands $\left\{ \Omega_{\beta}\right\} $
are indexed as $\beta=-\left(N-1\right),-\left(N-2\right),..,0,...,\left(N-2\right),\left(N-1\right)$.
Note that Eq.~\eqref{eq:autoCorr} shows that the  profile $f\left(r\right)$ is
simply the discrete autocorrelation of the sideband amplitudes. We
can thus make use of the convolution theorem to take the discrete
Fourier transform (DFT) of both sides:
\begin{equation}
f\left(k\right)=\left|\Omega_{k}\right|^{2},
\end{equation}
where we define the DFT as $f\left(k\right)=\sum_{r}e^{-ikr}f\left(r\right)$
and $\Omega_{k}=\sum_{\beta}e^{-ik\beta}\Omega_{\beta}$, introducing the inverse
DFT as   $f\left(r\right)=\frac{1}{\left(2N-1\right)}\sum_{k}e^{ikr}f\left(k\right)$
and $\Omega_{\beta}=\frac{1}{\left(2N-1\right)}\sum_{k}e^{ik\beta}\Omega_{k}$.
We know that any choice of $f\left(\abs{n-m}\right)$ which yields   a physically
valid dissipator must be positive semidefinite, and therefore $f\left(k\right)$
must be real and non-negative. We can take its square root
and get
\begin{equation}
\left|\Omega_{k}\right|=\sqrt{f\left(k\right)}.
\end{equation}
Now, we can look for solutions with $\Omega_{k}$   real, thus yielding
\begin{equation}\label{eq:sidebands}
\Omega_{\beta}=\frac{1}{\left(2N-1\right)}\sum_{k}e^{ik\beta}\sqrt{f\left(k\right)}
\end{equation}
which are the desired sideband amplitudes. One can numerically compute
these amplitudes using a fast Fourier transform and then check that
the amplitudes yield the desired  profile by computing
their autocorrelation (Eq.~\eqref{eq:autoCorr}). In Fig.~\ref{fig:Exp_SidebandConstruction}, we demonstrate this procedure for a long-range spatial profile $f\left(\abs{n-m}\right)=\frac{1}{\left(\left|n-m\right|+1\right)^{\alpha}}$ and a short-range spatial profile $f\left(\abs{n-m}\right)=e^{-|n-m|/\chi}$.

\section{Parameter estimates for cavity QED experiments}

For the construction to hold, we require that the detuning, $\Delta$, and cavity loss rate, $\gamma$, are sufficiently large. Specifically, we require that $\Delta\gg \mu N$, $\Delta \gg \omega_L - \omega_c - \omega_{g,n}$, and $\Delta \gg \Omega_M, g$. These conditions result in the excited atomic level $\ket{e}_n$ having approximately the same energy along the entire chain and only participating virtually in the dynamics. We also require that $\gamma \gg g \Omega_M / \Delta$ and $\gamma >> \delta_{\beta,m}$, which allow us to adiabatically eliminate the photon and suppress coherent spin-exchange interactions respectively. Lastly, we require that $\Delta^2 \gamma \gg g^2 \Omega_M^2 / \textrm{min}\abs{\omega_{g,n}-\omega_{g,m}}$ so that only the resonant process Eq.~\eqref{reson} takes place.

Below we provide estimates for the parameters involved in the experiment of Ref.~\cite{Monika} and show that our construction is accessible to current experimental systems. The number of lattice sites is  $N\approx$100 and it is comparable to the number of sidebands $\approx$~50-100 in the Raman beam. 
The Raman beam Rabi frequency $\left|\Omega\right|$ and cavity mode coupling $g$ range are on the order of a few MHz. The carrier frequency $\omega_{L}$ is $\approx$ 
384THz, while the sideband frequencies of the Raman beam satisfy $\max\left\{ \left|\tilde{\omega}_{\alpha}-\tilde{\omega}_{\beta}\right|\right\} \approx$ 10kHz. The Zeeman splitting $\omega_{g,n}=\mu n$ on each site is given by a magnetic field $\mu\cdot L\approx$~500kHz, where $L$ is the length of the cloud. Finally, the cavity decay  $\gamma$ is in the range $\approx$ 200kHz -- 10 of MHz. Per these experimental parameters, we can estimate that $\mu N/\Delta\sim 10^{-5}$, $(\omega_L-\omega_c-\omega_{g,n})/\Delta\sim10^{-4}$, $\Omega_M/\Delta \sim g/\Delta \sim 10^{-4}$, $\delta_{\beta,m}/\gamma\sim10^{-1}$, $g\Omega_M/(\Delta\gamma)\sim 10^{-4}$, and $g^2 \Omega_M^2 / (\textrm{min}\abs{\omega_{g,n}-\omega_{g,m}}\Delta^2 \gamma)\sim 10^{-1}$. The conditions for achieving the desired effective dynamics are thus satisfied.

While we should safely be able to ignore the coherent spin-exchange dynamics described in Eq.~\ref{eq:effectiveModel}, we can further examine the robustness of our platform to subleading effects. In Fig.~\ref{fig:Exp_LRHam}, we plot the correlation dynamics when both dissipative and coherent dynamics are present. As the relative strength of the coherent dynamics is increased, the pattern of correlations demonstrates a cross-over between the dissipative confinement pattern discussed in the main text and correlation fronts characteristic of a purely coherent dynamics arising from a spin-exchange Hamiltonian. The dissipative confinement pattern survives even when the coherent dynamics is one-fourth as strong as the dissipative dynamics; the experimental estimate of $\delta_{\beta,m}/\gamma\sim10^{-1}$ is well below this threshold.

The subleading spin-exchange Hamiltonian rotates the collective spin similar to a uniform external field and one may consider using it to tune the correlation pattern. However, as the spin-exchange dynamics also inherits a spatial profile, it complicates the spread of correlations and therefore makes an inconvenient tool to engineer a desired pattern arising from the dissipative dynamics. A uniform external field cannot create any spatial correlations, and therefore acts as a simple knob that tunes dynamics arising from the dissipator.

We note that such a uniform external field requires an addition to the experimental construction. It can be implemented, for example, by shining an additional multifrequency Raman beam that directly couples the two lower atomic levels with sideband frequencies corresponding to the change in energy splittings on each site. The amplitude $\omega_{F}$ characterizing the strength of this effective magnetic field would be set by the Rabi frequency $\Omega_x$ of this additional Raman beam, which typically takes on values between 1kHz to 50kHz in the experiment referenced above.

Lastly, we consider the effect of local losses from spontaneous Raman scattering of the individual atoms into free space. These incoherent spin losses are present in all experiments and introduce undesired decohering effects that deplete the total magnetization of the system. Following Ref.~\cite{estimate}, we estimate the number of spontaneous emission events after a time $t$ as $N_\Sigma\simeq N\Sigma t \Omega^2_M/\Delta^2$, with $\Sigma$ being the decay rate and $N$ being the total number of excitations in the system. The relevant time scale is set by the rate at which non-local dissipation generates correlations, given by $t\sim 1/\kappa\sim \gamma \Delta^2/g^2\Omega^2_M$. Correlations generated on this time scale are preserved if the number of spontaneous emission events is small compared to the number of atoms in the system: we require $N_\Sigma\ll N$. To satisfy this requirement, we need $\Sigma \gamma/g^2\ll1$, which together with the above stated condition of leaky cavity $\gamma\gg g\Omega_M/\Delta$, yields $\Sigma\ll (\Delta/\Omega_M)^2\gamma$. In the setup of Ref.~\cite{Monika},  we have $\Omega_M/\Delta\sim 10^{-4}$, and as both $\Sigma$ and $\gamma$ are typically of the order of a few MHz, correlations generated from non-local dissipation are protected against  scattering into free space for a long window of time. We can alternatively combine the conditions $\Sigma \gamma/g^2\ll1$ and $\gamma\gg g\Omega_M/\Delta$ into $g \gg \Sigma \Omega_M / \Delta$, which states that the coupling between the atoms and the cavity mode should be strong enough that coherent cavity-mediated transitions between atomic energy levels should be much more frequent than incoherent transitions due to spontaneous emission. For the parameters of Ref.~\cite{Monika},  $g$ and $\Sigma$ are of the order of a few MHz and $\Omega_M/\Delta\sim 10^{-4}$ so this strong coupling condition is satisfied.

\vspace{1cm}

\section{Finite wavevector squeezing parameter}

Here, we show that the finite wavevector squeezing parameter described in the main text acts as a witness of metrologically useful entanglement for the task of measuring a spatial Fourier component of an external field. Consider a spatially varying magnetic field $\vec{B}(\vec{r})=\left(0,0,B(\vec{r})\right)$ pointing in the $z$-direction of the coordinate system applied to a spin chain of $N$ spins represented by operators $\hat{\vec{S}}_{n}$, where $n$ indexes the position of each spin. If the field is applied for a time $t$, the effect on a system state $\rho_{0}$ is
\begin{equation}\label{eq:postsignalstate}
\rho_{0}\rightarrow\rho_{t}\left(\{B_{n}\}\right)=\hat{U}\left(t\right)\rho_{0}\hat{U}\left(t\right)^{\dagger}
\end{equation}
where $\hat{U}\left(t\right)=\exp\{i\hat{H}t\}$, $\hat{H}=\sum_{n=1}^{N}B_{n}\hat{S}^{z}_{n}$, and $B_{n}=B\left(\vec{r}_{n}\right)$. Letting $\vec{r}_n=na$ and setting the lattice constant as $a=1$, we can decompose the magnetic field in terms of Fourier components as $B_{n}=\frac{2}{N}\sum_{k}B_{k}\cos\left(k n\right)$ with $k\in\frac{2\pi}{N}\{1,N\}$. The Fourier components are given by the inverse transform $B_{k}=\sum_{n}B_{n}\cos\left(k n\right)$. Our goal is to estimate $B_{k^{*}}$ for a desired wavevector $k=k^{*}$ using $M$ measurements performed on the state $\rho_{t}$, given that the magnetic field strengths $\{B_{n}\}$ are unkown a priori. Estimating $B_{k^{*}}$ thus amounts to estimating a linear function
\begin{equation}
f\left(\boldsymbol{\alpha},\boldsymbol{\theta}\right)=\boldsymbol{\alpha}\cdot\boldsymbol{\theta}
\end{equation} 
of unknown parameters $\boldsymbol{\theta}=\{B_{n}\}$ and known coefficients $\boldsymbol{\alpha}=\{\cos\left(k^{*}n\right)\}$. The precision bounds, known as Cramer-Rao bounds, for such multiparameter estimation tasks were derived in Ref.~\cite{Eldredge}. Let $\Theta_{k^{*}}$ be the estimator of $B_{k^{*}}$, with a mean $\mathbb{E}\left[\Theta_{k^{*}}\right]$ and variance $\Delta\left(\Theta_{k^{*}}\right)^{2}$. The lower bound on the variance sets the maximum achievable precision of the estimator. If the initial state $\rho_{0}$ has no entanglement, then $\Delta\left(\Theta_{k^{*}}\right)^{2}\geq\Delta_{\textrm{SQL}}\left(\Theta_{k^{*}}\right)^{2}$ where
\begin{equation}
\Delta_{\textrm{SQL}}\left(\Theta_{k^{*}}\right)^{2} = \frac{N}{Mt^{2}}\frac{1}{2^{\delta_{k^{*}\neq 0}}}
\end{equation}
is known as the standard quantum limit (SQL). We have used the shorthand $\delta_{k^{*}\neq 0}=1$ if $k^{*}\neq0$ and zero otherwise. If we allow entanglement in the initial state $\rho_{0}$, then $\Delta\left(\Theta_{k^{*}}\right)^{2}\geq\Delta_{\textrm{HL}}\left(\Theta_{k^{*}}\right)^{2}$ where
\begin{equation}\label{eq:CR_HL}
\Delta_{\textrm{HL}}\left(\Theta_{k^{*}}\right)^{2} = \frac{1}{Mt^{2}}
\end{equation}
is known as the Heisenberg limit (HL). The key point is that $\Delta_{\textrm{SQL}}\left(\Theta_{k^{*}}\right)^{2}/\Delta_{\textrm{HL}}\left(\Theta_{k^{*}}\right)^{2}\propto N$, therefore entanglement allows a factor $N$ scaling improvement in the precision of the estimator. Importantly, as the generators $\{\hat{S}_{n}^{z}\}$ commute with each other, the Heisenberg limit can in principle be saturated with an optimal choice of initial state $\rho_{0}$ and measurement protocol.

In order to put the above expressions for the SQL and HL in a more familiar context, consider the quantity $\phi_{0}=\frac{1}{N}B_{k^{*}=0}$ and its estimator $\Phi_{0}$. This quantity corresponds to measuring the uniform component of the magnetic field and has precision limits
\begin{align}
\Delta_{\textrm{SQL}}\left(\Phi_{0}\right)^{2} &= \frac{1}{NMt^{2}},\\
\Delta_{\textrm{HL}}\left(\Phi_{0}\right)^{2} &= \frac{1}{N^{2}Mt^{2}},
\end{align}
which are the usual expressions for the single parameter estimation problem typically considered in quantum-enhanced metrology.

We can similarly gain intuition for the $k^{*}\neq0$ case by framing it as a single parameter estimation problem. Let us define the Fourier transformed spin operators as $\hat{\boldsymbol{S}}_{k}=\sum_{n}e^{-ikn}\hat{\boldsymbol{S}}_{n}$, with the inverse transform being $\hat{\boldsymbol{S}}_{n}=\frac{1}{N}\sum_{k}e^{ikn}\hat{\boldsymbol{S}}_{k}$. Then, we can write the Hamiltonian generating evolution due to the magnetic field as $\hat{H}=\frac{2}{N}\sum_{k}B_{k}\operatorname{Re}\{\hat{S}_{k}^{z}\}$ where $\operatorname{Re}\{\hat{\boldsymbol{S}}_{k}\}=\sum_{n}\cos\left(kn\right)\hat{\boldsymbol{S}}_{n}$. The variance of the estimator $\Theta_{k^{*}}$ is then bounded by generator of $B_{k^{*}}$ in $\hat{H}$~\cite{Pezze,Eldredge,Altenburg}. Specifically, we have $\Delta\left(\Theta_{k^{*}}\right)^{2}\geq\Delta_{\textrm{G}}\left(\Theta_{k^{*}}\right)^{2}$ where $\Delta_{\textrm{G}}\left(\Theta_{k^{*}}\right)^{2}=1/(Mt^{2}4\Delta\left(\hat{G}_{k^{*}}\right)^{2})$ and $\hat{G}_{k^{*}}=\partial\hat{H}/\partial B_{k^{*}}=\frac{2}{N}\operatorname{Re}\{\hat{S}_{k^{*}}^{z}\}$. This yields the precision bound
\begin{equation}\label{eq:CR_var}
\Delta_{\textrm{G}}\left(\Theta_{k^{*}}\right)^{2}=	\Delta_{\textrm{SQL}}\left(\Theta_{k^{*}}\right)^{2}\frac{2^{\delta_{k^{*}\neq 0}}}{16}\frac{N}{\Delta\left(\operatorname{Re}\{\hat{S}_{k^{*}}^{z}\}\right)^{2}}.
\end{equation}
The bound in Eq.~\eqref{eq:CR_var} is typically looser than the one in Eq.~\eqref{eq:CR_HL}, but provides intuition for how decreasing the variance of the operator $\hat{S}_{k^{*}}^{z}$ increases the potential precision in the estimate of $B_{k^{*}}$.

We now give an example of an estimator that can exploit squeezing of this variance to provide a metrological advantage. Consider an experiment that makes $M$ measurements of an observable $\hat{\mu}$ on the state $\rho_{t}(\theta)$ that depends on the unknown parameter of interest, $\theta$. Let the expected value of this observable be $\bar{\mu}=\textrm{Tr}\{\hat{\mu}\rho_{t}(\theta)\}\equiv f(\theta)$, where we have made the dependence on $\theta$ explicit, and the sample average of the measurements be $\bar{\mu}_M$. The method of moments (MOM) estimator is defined as the value of $\theta$ for which the expectation value $\bar{\mu}$ equals the sample average: $\Theta_{\textrm{mom}}=f^{-1}(\bar{\mu}_{M})$~\cite{Pezze}. We have $\bar{\mu}_{M}=f(\Theta_{\textrm{mom}})$, and in the limit of many measurements, the law of large numbers states that $\bar{\mu}_{M}\approx\bar{\mu}=f(\theta)$. Therefore, we expect $\Theta_{\textrm{mom}}\approx\theta$ and we can expand $\bar{\mu}_{M}=f(\Theta_{\textrm{mom}})$ around $\Theta_{\textrm{mom}}=\theta$:
\begin{equation}
f(\Theta_{\textrm{mom}}) \approx f(\theta) + \frac{\partial f}{\partial \theta}\Biggr\rvert_{\theta} \left(\Theta_{\textrm{mom}}-\theta \right).
\end{equation}
Plugging in $\bar{\mu}_M$ and $\bar{\mu}$, we have
\begin{equation}\label{eq:MOMexpansion_invert}
\bar{\mu}_{M} \approx \bar{\mu}\left(\theta\right) + \frac{\partial \bar{\mu}}{\partial \theta}\Biggr\rvert_{\theta} \left(\Theta_{\textrm{mom}}-\theta \right).
\end{equation}
The convenient aspect of the MOM estimator is that the value $\Theta_{\textrm{mom}}$ can be extracted directly from the experimental sample average $\bar{\mu}_{M}$. Extracting the estimate, however, requires knowledge of the functional form $\bar{\mu}(\theta)$, or its inverse, as is the case in the usual Ramsey metrology protocol where typically $\bar{\mu}(\theta)\propto\cos^{2}(\theta)$.  Alternatively, the MOM estimate can be extracted using the above Taylor expansion if we know a calibration value $\theta_{c}$ that is close to the true unknown value $\theta$, as well as $\bar{\mu}(\theta_{c})$ and $\frac{\partial \bar{\mu}}{\partial \theta}\rvert_{\theta_{c}}$ with high precision~\cite{Altenburg}:
\begin{equation}\label{eq:MOMexpansion_calib}
\bar{\mu}_{M} \approx \bar{\mu}\left(\theta_{c}\right) + \frac{\partial \bar{\mu}}{\partial \theta}\Biggr\rvert_{\theta_{c}} \left(\Theta_{\textrm{mom}}-\theta_{c} \right).
\end{equation}
In either case, the variance of the estimator can be calculated in the limit of large $M$ by identifying $\Delta\left(\Theta_{\textrm{mom}}\right)^{2}\approx\left(\Theta_{\textrm{mom}}-\theta \right)^2$ and $\Delta\left(\bar{\mu}(\Theta_{\textrm{mom}})\right)^{2}\approx M \left(\bar{\mu}(\Theta_{\textrm{mom}})-\bar{\mu}(\theta) \right)^2$ in Eq.~\eqref{eq:MOMexpansion_invert}, or by replacing $\theta\rightarrow\theta_{c}$ in these expressions and then using them with Eq.~\eqref{eq:MOMexpansion_calib}. Letting $\hat{G}_{\theta}=\partial\hat{H}/\partial \theta$, we have $\frac{\partial \bar{\mu}}{\partial \theta}\rvert_{\theta_{c}}=-it\textrm{Tr}\{\left[\hat{G}_{\theta},\hat{\mu}\right]\rho_{t}(\theta)\}$. The variance of the MOM estimator is then
\begin{equation}
\Delta\left(\Theta_{\textrm{mom}}\right)^{2} = \frac{\langle \Delta\left(\hat{\mu}\right)^{2}\rangle}{Mt^{2}\lvert\langle \left[\hat{G}_{\theta},\hat{\mu}\right] \rangle\rvert^{2}}
\end{equation}
where $\langle \cdot \rangle = \textrm{Tr}\{\cdot \rho_{t}(\theta)\}$. For our problem to estimate $\theta=B_{k^{*}}$, we have $\hat{G}_{\theta}=\frac{2}{N}\operatorname{Re}\{\hat{S}_{k^{*}}^{z}\}$. We pick a measurement observable $\hat\mu=\boldsymbol{e}_{\perp}\cdot\operatorname{Re}\{\hat{\boldsymbol{S}}_{k^{*}}\}$ where $\boldsymbol{e}_{\perp}$ represents a unit vector in the plane perpendicular to both the mean spin direction, $\boldsymbol{e}_{s}$, and the direction of the external magnetic field, $\boldsymbol{e}_{z}$. The variance of our MOM estimate for $B_{k^{*}}$ is
\begin{align}
\Delta_{\textrm{mom}}\left(\Theta_{k^{*}}\right)^{2}&=\frac{N}{Mt^{2}}\frac{N\langle \Delta\left(\boldsymbol{e}_{\perp}\cdot\operatorname{Re}\{\hat{\boldsymbol{S}}_{k^{*}}\}\right)^{2}\rangle}{\lvert\langle \boldsymbol{e}_{s}\cdot\hat{\boldsymbol{S}}_\textrm{tot} \rangle+\langle \boldsymbol{e}_{s}\cdot\operatorname{Re}\{\hat{\boldsymbol{S}}_{2k^{*}}\} \rangle\rvert^{2}}\\
&=\frac{N}{Mt^{2}}\frac{N\langle \Delta\left(\boldsymbol{e}_{\perp}\cdot\operatorname{Re}\{\hat{\boldsymbol{S}}_{k^{*}}\}\right)^{2}\rangle}{4\lvert\sum_{n}\cos^{2}\left(k^{*}n\right)\langle \boldsymbol{e}_{s}\cdot\hat{\boldsymbol{S}}_{n} \rangle\rvert^{2}}\\
&\leq \frac{N}{Mt^{2}}\frac{N\langle \Delta\left(\boldsymbol{e}_{\perp}\cdot\operatorname{Re}\{\hat{\boldsymbol{S}}_{k^{*}}\}\right)^{2}\rangle}{4\lvert\langle \boldsymbol{e}_{s}\cdot\hat{\boldsymbol{S}}_\textrm{tot} \rangle\rvert^{2}}
\end{align}
where $\hat{\boldsymbol{S}}_\textrm{tot} = \hat{\boldsymbol{S}}_{k=0} = \sum_{n}\hat{\boldsymbol{S}}_{n}$ and we have used the fact that spin operators satisfy the commutation relation $\left[\hat{S}_{k}^{\alpha},\hat{S}_{k'}^{\beta}\right]=i\varepsilon^{\alpha\beta\gamma}\hat{S}_{k+k'}^{\gamma}$ for $\alpha,\beta,\gamma\in\{x,y,z\}$. We can thus upper bound the variance of the MOM estimator as 
\begin{equation}
\Delta_{\textrm{mom}}^\textrm{ub}\left(\Theta_{k^{*}}\right)^{2}=\Delta_{\textrm{SQL}}\left(\Theta_{k^{*}}\right)^{2} \xi_{k^{*}}^{2}
\end{equation}
with
\begin{equation}\label{eq:kWineland}
\xi_{k^{*}}^{2} = \frac{2^{\delta_{k^{*}\neq 0}}N\langle \Delta\left(\boldsymbol{e}_{\perp}\cdot\operatorname{Re}\{\hat{\boldsymbol{S}}_{k^{*}}\}\right)^{2}\rangle}{\lvert\langle \boldsymbol{e}_{s}\cdot\hat{\boldsymbol{S}}_\textrm{tot} \rangle\rvert^{2}},
\end{equation}
For $k^{*}=0$, we get Wineland's metrological squeezing parameter~\cite{Pezze}, and therefore  Eq.~\eqref{eq:kWineland} serves as a generalization of Wineland's parameter to the case of sensing specific Fourier components of a spatially varying field. If $\xi_{k^{*}}^{2}<1$, then a MOM estimator can exploit entanglement in the system to measure this Fourier component with a precision beyond the standard quantum limit. The estimator requires that we can measure $\langle \boldsymbol{e}_{\perp}\cdot\operatorname{Re}\{\hat{\boldsymbol{S}}_{k^{*}}\}\rangle$. Using the Fourier decomposition of $\hat{\boldsymbol{S}}_{k^{*}}$, this requires computing $\langle \boldsymbol{e}_{\perp}\cdot\hat{\boldsymbol{S}}_{n}\rangle$, which can easily be extracted from simultaneous projective measurements of $\langle \hat{S}_{n}^{z} \rangle$ of each spin after rotation of $\boldsymbol{e}_{\perp}$ to the $\boldsymbol{e}_{z}$ basis. Such measurements are routinely performed in cold atom systems using flourescence imaging to determine the occupation of the atoms in each of their internal states (corresponding to spin up and spin down).

In general, if we are trying to characterize the metrological utility of a state $\rho_{0}$ with a fixed mean spin direction $\boldsymbol{e}_{s}$, we can presume control over the direction of the external magnetic field $\vec{B}(na)$ that is being sensed and align it for greatest sensitivity. The metrological gain to sense a Fourier component at wavevector $k$ of the field can then be quantified via the finite wavevector squeezing parameter
\begin{equation}\label{eq:kSqueezing}
\left(\xi_{k}^{(\textrm{W})}\right)^{2} = \text{\ensuremath{\min}}_{\boldsymbol{e}_{\perp}}\Biggr\{\frac{2^{\delta_{k\neq 0}}N\langle \Delta\left(\boldsymbol{e}_{\perp}\cdot\operatorname{Re}\{\hat{\boldsymbol{S}}_{k}\}\right)^{2}\rangle}{\lvert\langle \boldsymbol{e}_{s}\cdot\hat{\boldsymbol{S}}_\textrm{tot} \rangle\rvert^{2}}\Biggr\},
\end{equation}
where the minimization is performed over all directions $\boldsymbol{e}_{\perp}$ that are perpendicular to the mean spin direction $\boldsymbol{e}_{s}$. One should be careful in the interpretation of the above generalized squeezing parameter. The spin operators at a given wavevector $k$ do not form a closed spin algebra, and therefore there is no single Bloch sphere that can be associated with this spin mode. Therefore, the usual intuition of squeezing a quadrature of the spin on its Bloch sphere does not hold. Nonetheless, Eq.~\eqref{eq:kSqueezing} does quantify the amount of metrological gain that can be achieved due to entanglement in the state using projective measurements of local spin operators (e.g. $\langle \hat{S}_{n}^{z} \rangle$). The reason is that reducing the variance of the generator of the estimated parameter helps increase the precision of the estimate, as described in Eq.~\eqref{eq:CR_var}. In general, the precision of this MOM estimator is worse than the optimal precision set by Eq.~\eqref{eq:CR_HL}, but one may hope to find squeezing values, $(\xi_{k}^{(\textrm{W})})^{2}$, which scale as $1/N$ and therefore provide the same scaling advantage. More optimal estimators can saturate the bound of Eq.~\eqref{eq:CR_HL}~\cite{Eldredge, KQian,TQian,Sekatski,Wolk}.



\end{document}